\documentstyle[12pt]{article}

\def\beq{\begin{equation}} \def\eeq{\end{equation}}
\def\bea{\begin{eqnarray}} \def\eea{\end{eqnarray}}

\def\beann{\begin{eqnarray*}} \def\eeann{\end{eqnarray*}}

\let\a=\alpha \let\be=\beta \let\g=\gamma \let\de=\delta
\let\e=\varepsilon   
\let\dh=\vartheta \let\k=\kappa \let\la=\lambda \let\m=\mu
\let\n=\nu  \let\p=\pi \let\r=\rho \let\s=\sigma
 
\let\ph=\varphi

\let\qd=\quad  \def\qqqd{\qquad\qquad}

\let\Ts=\textstyle \let\Ds=\displaystyle 

\def\0{\over } \def\1{\vec }     \def\2{{1\over2}} \def\4{{1\over4}}
\def\5{\bar }  \def\6{\partial } \def\7#1{{#1}\llap{/}}

\def\<{\langle } \def\>{\rangle }

\let\auf=\uparrow \let\ab=\downarrow

\def\i{{\rm i}} \def\tr{\mbox{tr}}

\def\ctg{\mbox{\,ctg}}
\def\sh{\mbox{\,sh}} 
 
\def\cth{\mbox{\,cth}}
  
\def\sign{\mbox{\,sign}}

\begin{document}

\thispagestyle{empty}
\begin{center}
{\Large {\bf The Yangian Symmetry of the Hubbard Models with
Variable Range Hopping\\}}
\vspace{2cm}
{\large Frank G\"{o}hmann$^\dagger$
and Vladimir Inozemtsev$^\ddagger$}\\
\vspace{5mm}
$^\dagger$Department of Physics, Faculty of Science, University
of Tokyo,
Hongo 7-3-1, Bunkyo-ku, Tokyo 113, Japan
\footnote[1]{e-mail: frank@monet.phys.s.u-tokyo.ac.jp}\\
$^\ddagger$ISSP, University of Tokyo, 7-2-21 Roppongi, Minato-ku,
Tokyo 106, Japan
\vspace{15mm}

{\large {\bf Abstract}}
\end{center}
\begin{list}{}{\addtolength{\rightmargin}{10mm}
               \addtolength{\topsep}{-5mm}}
\item
We present two pairs of Y($sl_2$) Yangian symmetries for the
trigonometric and hyperbolic versions of the Hubbard model with
non-nearest-neighbour hopping. In both cases the Yangians are mutually
commuting, hence can be combined into a Y($sl_2$)$\oplus$Y($sl_2$)
Yangian. Their mutual commutativity is of dynamical
origin. The known Yangians of the Haldane-Shastry spin chain and
the nearest neighbour Hubbard model are contained as limiting
cases of our new representations.
\end{list}

\vspace{2cm}

Yangian quantum groups were introduced by Drinfel'd
\cite{Drinfeld85} more then ten years ago. His original
intention was to put the algebraic structure underlying the
Yang-Baxter exchange relations with rational $R$-matrix,
responsible for the integrability of the most prominent
integrable systems, into the mathematically more conventional
context of Hopf algebras. Yangian quantum groups and their
representation theory are thus intimately connected with
the classification of integrable quantum systems.

However, during the past few years it became apparent that Yangians
can also play a physically very interesting role as additional
symmetries of integrable systems, and moreover, that Yangians are
part of the symmetry algebra of such well studied integrable systems
as the nearest neighbour Heisenberg model \cite{Bernard93}, or the
nearest neighbour Hubbard model \cite{UgKo94}, if considered on
infinite lattices.  These symmetries have been overlooked for many
years, since it was unusual to deal directly with infinite systems.
Instead all conventional approaches to integrable systems like the
various Bethe Ans\"atze start from finite systems, usually under
periodic boundary conditions, and the thermodynamic limit is only
performed at a later stage of calculation. The Yangian symmetries
of the nearest neighbour Hubbard and Heisenberg models are
incompatible with periodic boundary conditions. For this reason
they do not combine with Bethe Ansatz methods.

Now there are two recent developments
that make a utilisation of Yangians for integrable systems
feasible.  First, methods have been developed to deal directly
with infinite systems. Within this so-called symmetry based
approach it became possible to calculate for instance higher
order spin correlators for the XXZ-chain \cite{JMMN92}. Quantum
groups play an essential role here. Second, interesting
integrable systems with a finite number of degrees of freedom have
been discovered which exhibit Yangian symmetry compatible with
periodic boundary conditions, most prominent among these the
Haldane-Shastry spin chain \cite{Haldane88,Shastry88}. Until the
discovery of its Yangian symmetry \cite{HHTBP92} the high
degeneracy of its spectrum remained a puzzle, which is now
resolved by Yangian representation theory \cite{BGHP93}.

Below we present a new pair of mutually commuting
representations of the Y($sl_2$) Yangian in terms of
Fermi operators, which form a Yangian symmetry of the Hubbard model
with non-nearest-neighbour hopping of Gebhard and Ruckenstein
\cite{GeRu92,BaGe95}. The model itself contains the usual
Hubbard model as well as the Haldane-Shastry spin chain as
certain limiting cases, and so do the generators of its Yangian
symmetry.

The model describes itinerant electrons of spin $\s$ created by
$c_{j\s}^+$ at site $j$ of a one dimensional lattice. The
probability amplitude for hopping between sites $j,k$ will be
denoted by $t_{jk}$. Two electrons of different spin encountering
each other on the same lattice site feel an on-site repulsion
$U > 0$.
\beq \label{ham}
     H = \sum_{j, k} t_{jk} c_{j \s}^+ c_{k \s} +
         2U \sum_j (c_{j \auf}^+ c_{j \auf} - {\Ts \frac{1}{2}})
                   (c_{j \ab}^+ c_{j \ab} - {\Ts \frac{1}{2}}) \qd.
\eeq
The constants here have been chosen for later convenience, and
$t_{jj} := 0$ by definition. Throughout this letter we are using
sum convention with respect to Greek indices. We will consider
two different choices of translational invariant hopping
amplitudes, $t_{jk} = t_{j - k}$. For $n \ne 0$ let
\beq \label{hop}
     t_{n} := - \i \sh(\k) \sh^{-1}(\k n) \qd,\qd n \ne 0 \qd.
\eeq
Then our choices are given by $\k = \i\, \p/N$ for a finite
lattice of $N$ sites, and $\k > 0$ for an infinite lattice. The
energy scale has been chosen such as to give
hopping amplitudes of absolute value 1 between neighbouring
sites. The summation indices run from $0$ to $N - 1$ in the
trigonometric case, and over all integers in the hyperbolic case. The
thermodynamic limit of the trigonometric model and the limit
$\k \rightarrow 0$ of the hyperbolic model coincide. In both
cases $t_{jk}$ turns into $-\i/(j - k)$. This is the true $1/r$
Hubbard model\footnote{It may be interesting to notice, that
the trigonometric and hyperbolic hopping amplitudes above can be
interpreted as $q$-deformed $1/r$-hopping. The notion of
$q$-deformation is defined by $r_q := (q^r - q^{-r})/(q - q^{-1})$.
Setting $q$ equal to $e^{\k}$ the hopping amplitudes become
$t_{jk} = - \i/(j - k)_q$. The trigonometric case corresponds to
$q$ being the $N$-th root of unity, the hyperbolic case to $q > 1$.}.
In the limit $\k \rightarrow \infty$, the hyperbolic model turns,
up to a canonical transformation, which is described below, into
the nearest neighbour Hubbard model.

In order to understand the physical meaning of the above kind of
hopping amplitudes, one has to consider the dispersion relation
of the free model ($U=0$) \cite{GeRu92,BaGe95}. In the
trigonometric case we obtain
\beq \label{dispt}
     \e(p) = \sum_{n=1}^{N-1} t_n e^{\i pn} = \frac{N}{\p}
             \sin \left( \frac{\p}{N} \right)(\p - p) \qd,
\eeq
where  $p = 2\p(m + 1/2)/N$,
$m = 0, \dots , N - 1$. In the thermodynamic limit this yields
$\e(p) \rightarrow \p - p$. The dispersion relation
(\ref{dispt}) is linear in the first Brillouin zone, the model
is chiral. It contains only left moving particles. The physically
most interesting point about this chiral model is the appearance
of a Mott transition at finite $U > 0$ \cite{GeRu92,GGR94}. In the
hyperbolic case the dispersion relation is
\beq
     \e(p) = \sum_{n=1}^\infty t_n e^{\i pn} =
             2 \sh(\k) \sum_{n=1}^\infty \frac{\sin(pn)}{\sh(\k n)}
             \qd.
\eeq
The last expression is easily recognised as being, up to a
redefinition of scales, the logarithmic derivative of the Jacobi
theta function $\dh_4$. As a function of $\k$ it interpolates
between the sinusoidal dispersion relation of the of the nearest
neighbour model and the saw tooth shaped dispersion relation of
the $1/r$ model.

The local U(1) transformation $c_{j\s} \rightarrow e^{\i \ph_j}
c_{j\s}$, $\ph_j$ real, does not alter the canonical
anticommutation relations between the Fermi operators. The local
electron densities $c_{j\s}^+ c_{j\s}$ are invariant under this
transformation, hence the interaction part of the
Hamiltonian (\ref{ham}) is as well. This means that we can always use a U(1)
transformation to modify the hopping term to our convenience. The
modified model will be completely equivalent to the original
one. Consider the case $\ph_j = j \p$. This transformation
introduces a factor of $(-1)^{j-k}$ into the expression for the
hopping amplitudes and shifts the dispersion relations by a half
period. Using this transformation our conventions meet the
conventions of Gebhard and Ruckenstein \cite{GeRu92}. To recover
the nearest neighbour Hubbard model in its familiar form,
we do not only have to consider $\k \rightarrow \infty$, but in
addition the above transformation with $\ph_j = j \p/2$.
This transformation removes the factor of ``$\i$'' in front of
the hopping amplitude, changes the hopping amplitude to an even
function, and shifts the dispersion relation by a quarter period.
Hence the quadratic bottom of the sinusoidal band is shifted to
$p=0$.

There is yet another important canonical transformation, namely
\beq \label{cano}
     c_{j \ab} \rightarrow c_{j \ab} \qd, \qd
     c_{j \auf} \rightarrow c_{j \auf}^+ \qd, \qd
     U \rightarrow - U \qd.
\eeq
This transformation leaves every Hamiltonian of the form (1) with
antisymmetric hopping matrix invariant, but the global spin
operators and, in our case, the Yangian generators (see below) are not.
It is responsible for the doubling of the Yangian.

The natural language for writing down the $sl_2$ generators of
the rotational symmetry of the Hamiltonian (\ref{ham}) is, of
course, in terms of spin operators, which are linear combinations of
products of one creation and one annihilation operator at the
same site. For the formulation of our Yangian generators below
it turns out to be useful to extend this concept to spin-like
operators with indices corresponding to different sites. We
arrange the pair of operators $c_{j\s}^+ c_{k\tau}$ in a
$2 \times 2$-matrix labeled by spin indices $\s$, $\tau$ in the
usual tensor product convention, $(S_{jk})_{\tau}^{\s} :=
c_{j\s}^+ c_{k\tau}$, and then set
\beq
     S_{jk}^\a := \tr(\bar{\s}^\a S_{jk}) \qd, \qd
     S_{jk}^0 := \tr(S_{jk}) \qd, \qd S_j^\a := S_{jj}^\a \qd, \qd
     S_j^0 := S_{jj}^0 \qd,
\eeq
where the $\s^\a$ are the Pauli matrices, and the bar denotes
complex conjugation. Our definition implies $(S_{jk}^\a)^+
= S_{kj}^\a$, $(S_{jk}^0)^+ = S_{kj}^0$. $\frac{1}{2}S_j^\a$
and $S_j^0$ are the spin density and electron density operators,
respectively. The algebra of the operators $S_{jk}^\a$ is rather
rich. We obtain the commutators
\bea \label{c1}
     [S_{jk}^0,S_{lm}^0] & = &
         \de_{kl} S_{jm}^0 - \de_{mj} S_{lk}^0 \qd, \\[1ex]
     \label{c2}
     [S_{jk}^0,S_{lm}^\a] & = &
         \de_{kl} S_{jm}^\a - \de_{mj} S_{lk}^\a \qd, \\[1ex]
     \label{c3}
     [S_{jk}^\a,S_{lm}^\be] & = &
         \de^{\a \be} \left(
         \de_{kl} S_{jm}^0 - \de_{mj} S_{lk}^0 \right)
         + \i \, \e^{\a \be \g} \left(
         \de_{kl} S_{jm}^\g + \de_{mj} S_{lk}^\g \right) \qd.
\eea
However, there are other relations. For the construction
of our Yangian generator and the verification of the Yangian Serre
relations below, we further need the following,
\bea \label{o1}
     S_{jk}^\a S_{lm}^\a
         +  S_{jk}^0 S_{lm}^0 + 2 S_{jm}^0 S_{lk}^0 & = &
         4 \de_{kl} S_{jm}^0 + 2 \de_{lm} S_{jk}^0
         \qd, \\[1ex] \label{o2}
     S_{jk}^0 S_{lm}^\a + S_{lm}^0 S_{jk}^\a +
         S_{lk}^0 S_{jm}^\a + S_{jm}^0 S_{lk}^\a & = &
         \de_{jk} S_{lm}^\a + \de_{lm} S_{jk}^\a +
         \de_{lk} S_{jm}^\a + \de_{jm} S_{lk}^\a
         \qd, \\[1ex] \label{o3}
     S_{jk}^\a S_{lm}^\be + S_{jk}^\be S_{lm}^\a +
     S_{jm}^\a S_{lk}^\be + S_{jm}^\be S_{lk}^\a & = &
         \de^{\a \be} \left(
         S_{jm}^0 (2\de_{lk} - S_{lk}^0) + S_{jm}^\g S_{lk}^\g
         \right) \qd, \\[1ex] \label{o4}
     - \i \e^{\a \be \g} S_{jk}^\be S_{lm}^\g
         - S_{jm}^0 S_{lk}^\a + S_{lk}^0 S_{jm}^\a & = &
         2 \de_{lk} S_{jm}^\a + \de_{jk} S_{lm}^\a -
         \de_{lm} S_{jk}^\a \qd.
\eea
These relations generate a long list of succeedingly less general
relations by systematically equating all possible combinations of
site indices. Setting $j = k$ and $l = m$ in eq.\ (\ref{c3}), for
example, implies that the operators $\frac{1}{2}S_j^\a$ are spin
density operators,
\beq \label{co4}
     [S_j^\a,S_k^\be] = \de_{jk} 2\i \, \e^{\a \be \g} S_j^\g \qd.
\eeq
The Hamiltonian (\ref{ham}) now assumes the following form
\beq \label{eggs}
     H = \sum_{j,k} t_{jk} S_{jk}^0 + U \sum_j \left(
         (S_j^0 - 1)^2 - \Ts{\frac{1}{2}} \right) \qd.
\eeq
Since the particle number $I^0 = \sum_j S_j^0$ is
conserved, only the term $(S_j^0)^2$ is relevant in the
interaction part of the Hamiltonian. The other terms can be
removed by a shift of the chemical potential. We retained them
here to make obvious the invariance of $H$ under the
transformation (\ref{cano}). The operators of the total spin
are $I^\a := \frac{1}{2} \sum_j S_j^\a$. It follows from
(\ref{c1}) that they commute with the Hamiltonian. Their
$sl_2$ commutation relations are obtained by summing (\ref{co4})
over $j$ and $k$,
\beq \label{sl2}
     [I^\a,I^\be] = \i \, \e^{\a \be \g} I^\g \qd.
\eeq
Now everything is prepared to formulate our main result.
Consider the Hamiltonian (\ref{eggs}) with yet unspecified
antisymmetric hopping matrix, $t_{jk} = - t_{kj}$. Let
\beq
     J^\a := {\Ts \frac{1}{2}} \sum_{j,k} \left(
             (f_{jk} + h_{jk}(S_j^0 + S_k^0 - 2))S_{jk}^\a
             + g_{jk} \, \e^{\a \be \g} S_j^\be S_k^\g \right) \qd,
\eeq
where $g_{jk}$ and $h_{jk}$ are odd functions, and $f_{jj} =
g_{jj} = h_{jj} = 0$ by convention. Then $H$ commutes with
$J^\a$ if and only if the following functional equations between
the coefficients are satisfied,
\bea \label{f1}
     (g_{jl} - g_{kl}) h_{jk} & = & {\Ts \frac{\i}{2}} h_{jl} h_{kl}
       \qd, \qd j \ne k \ne l \ne j \qd, \\[1ex] \label{f2}
      \i U f_{jk}/2h_0 + g_{jk} h_{jk} & = & - {\Ts \frac{\i}{4}}
       \sum_l h_{jl} h_{kl}
       \qd, \qd j \ne k \qd, \\[1ex] \label{f3}
     \sum_l (f_{jl} h_{kl} - f_{kl} h_{jl}) & = & 0
       \qd, \\[1ex] \label{f4}
     t_{jk} & = & h_0 h_{jk} \qd.
\eea
Here $h_0$ is a free parameter which fixes the scale for $J^\a$.
The only solutions to these equations correspond to the cases of
trigonometric and hyperbolic hopping amplitudes (\ref{hop})
under consideration. In the trigonometric case we find
\beq
     f_{jk} = 0 \qd, \qd g_{jk} = {\Ts \frac{1}{2}}
         \ctg (\p (j - k)/N) \qd, \qd
     h_{jk} = \i \sin ^{-1} (\p (j - k)/N) \qd,
\eeq
whereas in the hyperbolic case
\beq
     f_{jk} = \frac{\sh(\k)(j - k)}{U \sh(\k(j - k))} \qd, \qd
     g_{jk} = {\Ts \frac{1}{2}} \cth(\k(j - k)) \qd, \qd
     h_{jk} = \i \sh^{-1} (\k(j - k)) \qd.
\eeq
$h_0$ has to be real in order for $J^\a$ to be selfadjoint. We
choose $h_0 = - \sin(\p/N)$ in the trigonometric case and
$h_0 = - \sh(\k)$ in the hyperbolic case. It is an unexpected fact
that $J^\a$ does not depend on $U$ in the trigonometric case.
Hopping part and interaction part of the Hamiltonian commute
separately with $J^\a$.

One easily checks that our conserved operator $J^\a$ turns into
known generators of Yangians in various limiting cases. In the
nearest neighbour Hubbard limit of the hyperbolic model
($\k \rightarrow \infty$) we obtain $f_{jk} \rightarrow
\de_{|j - k|,1}/U$, $g_{jk} \rightarrow \sign(j - k)/2$,
$h_{jk} \rightarrow 0$. After a canonical transformation
$c_{j\s} \rightarrow \i^j c_{j\s}$ we recover the Yangian generator
of Uglov and Korepin \cite{UgKo94}. In the limit $U \rightarrow
\infty$ at less than half filling the model reduces to the ``$t$-0''
chain \cite{WZC93} with all states with double occupancies
projected out from the Hilbert space. Because at half filling
hopping is not allowed anymore in this limit, one can set
$S_j^0 = 1$, and recover the Yangian generator of the
Haldane-Shastry chain \cite{HHTBP92} or its hyperbolic counterpart
to leading order in $(t/U)^2$.

Indeed $J^\a$ itself generates a representation of a Y($sl_2$)
Yangian. This is our second result. The spin operators
$I^\a$ and the conserved quantities $J^\a$ satisfy the relations
\beq \label{y1}
     \hspace{-90mm} [I^{\la},J^{\m}] = c_{\la \m \n} J^{\n} \qd,
\eeq
\beq
     \label{y2}
     \hspace{-2mm}
     [[J^{\la},J^{\m}],[I^{\r},J^{\s}]] +
     [[J^{\r},J^{\s}],[I^{\la},J^{\m}]] =
     - 4 \de (a_{\la \m \n \a \be \g} c_{\r \s \n} +
     a_{\r \s \n \a \be \g} c_{\la \m \n}) \{I^{\a},I^{\be},J^{\g}\},
\eeq
where $\de = - 1$ in the trigonometric case, $\de = 1$ in the
hyperbolic case, and the further abbreviations
\bea
     c_{\la \m \n} & := & \i \, \e^{\la \m \n} \\[1ex]
     a_{\la \m \n \a \be \g} & := & c_{\la \a \r} c_{\m \be \s}
                                    c_{\n \g \tau} c_{\r \s \tau}
                                    \qd, \\[1ex]
     \{x_1,x_2,x_3\} & := & {\Ts\frac{1}{6}}
                            \sum_{i \ne j \ne k \ne i} x_i x_j x_k \qd
\eea
have been used.
Eqs.\ (\ref{y1}) and (\ref{y2}) together with the defining
relation (\ref{sl2}) of the $sl_2$ Lie algebra are Drinfel'd's
definition of the Y($sl_2$) Yangian \cite{Drinfeld85}. Equation
(\ref{y1}) says that the $J^\a$ transform like a vector
representation of $sl_2$, and is easily confirmed for our $J^\a$.
Since both equations (\ref{y1}) and (\ref{y2}) are homogeneous,
we could have introduced a deformation parameter $h^2$ on the
right hand side of the Yangian Serre relation (25). Since this
parameter merely fixes the scale of $J^\a$ and has no deeper
physical meaning, we suppressed it here. We have confirmed
(\ref{y1}) by direct calculation. The calculation is lengthy.
Before we comment on it we formulate our third result.

Under the transformation (\ref{cano}) the generators $I^\a$,
$J^\be$ transform into an independent set of generators $I'^\a$,
$J'^\be$ of another representation of the Y($sl_2$) Yangian. The
two representations commute, hence can be combined to a
Y($sl_2$)$\oplus$Y($sl_2$) double Yangian. Their commutativity
is non trivial and is of dynamical origin, i.e.\ it relies on the
functional equations (\ref{f1}) - (\ref{f4}) between the
coefficients that define $J^\a$, $J'^\be$.

To check the Yangian Serre relation, the original
formulation  (\ref{y2}) is rather inappropriate. We used the
following simplification in the $sl_{2}$ case instead. Let
\beq \label{defk}
     K^\a := - \i \e^{\a \be \g} [J^\be,J^\g]
             - 4 \de (I^\be)^2 I^\a \qd.
\eeq
Then a short but slightly tricky calculation shows that
(\ref{y2}) is equivalent to the equation
\beq \label{2com}
     [J^\a,K^\be] + [J^\be,K^\a] = 0 \qd.
\eeq
The left-hand side of (30) has a property that turns out to be
very useful in practical calculations. It is traceless.
Assume we are given an operator $J^\a$, and we do already know
that it transforms as a vector representation of $sl_2$. Then
this knowledge assures the identity $[J^\a,K^\a] = 0$. It is
therefore sufficient to show that the left-hand side of equation
(\ref{2com}) is proportional to $\de^{\a \be}$. This is a severe
simplification, since the symmetrisation of the commutator
produces a lot of terms proportional to $\de^{\a \be}$, which
can be neglected according to the above argument. The explicit
expression for $K^\a$ in our case is
\beq
\begin{array}{l}
     K^\a = \frac{1}{2}
            {\Ds \sum_{j,k,l}} \left\{ (8g_{jk}g_{jl} - \de)
            S_j^\be S_k^\be S_l^\a + 2 A_{jl} A_{lk} S_{jk}^\a
            - 4\i A_{jk} (g_{jl} - g_{kl}) S_{jk}^0 S_l^\a \right.
            \\[1ex] \qqqd
            \left. + 2 A_{jk} (g_{jl} + g_{kl})
            \e^{\a \be \g} S_{jk}^\be S_l^\g
            + \i h_{jl} \e^{\a \be \g}
            (A_{jk} S_{jk}^\be - A_{kj} S_{kj}^\be)
            (S_{jl}^\g - S_{lj}^\g) \right\} ,
\end{array}
\eeq
where $A_{jk} := f_{jk} +
h_{jk}(S_j^0 + S_k^0 - 2)$. To verify (\ref{2com}), we
used the following relations among the coefficients $f_{jk}$,
$g_{jk}$, $h_{jk}$ in addition to their defining functional
equations above.
\bea
     f_{jk}(g_{jl} - g_{kl}) & = & {\Ts \frac{\i}{2}}
           (f_{jl}h_{kl} - f_{kl}h_{jl}) \qd,
           \qd j \ne k \ne l \ne j \qd, \\[1ex]
     g_{jk}g_{jl} + g_{kl}g_{kj} + g_{lj}g_{lk} & = &
     \de/4 \qd, \qd j \ne k \ne l \ne j \qd, \\[1ex]
     4 g_{jk}^2 + h_{jk}^2 & = & \de \qd, \qd j \ne k \qd.
\eea

The homogeneity of the lattices
has not been used in the verification of the Yangian Serre relation
in the bulk. However, it is necessary to guarantee the commutativity of
$J^\a$ with the Hamiltonian. This situation is similar to the
case of the Yangian symmetric spin chains. Therefore we
conjecture the existence of a Yangian symmetric long range
Hubbard Hamiltonian on an inhomogeneous lattice. In analogy to
the spin chain case \cite{Hikami95b} the generator of its
Yangian symmetry might be constructed by adding ``potential terms''
to the second order Yangian generator $K^\a$, eq.\ (\ref{defk}).

At this point we would like to emphasize that Yangian symmetry
does not imply integrability. Nevertheless, we strongly believe
that the models considered here are integrable, and are special
cases of a more general integrable non Yangian symmetric model
with elliptic hopping amplitudes. The proof of integrability
would provide the basis for an understanding of the
Haldane-Shastry chain and the nearest neighbour Hubbard model on
a common ground. At the present state of knowledge these models
appear rather unlike. The integrability of the Haldane-Shastry
chain has been shown by exploiting a mapping to a related
dynamical model \cite{TaHa94}, whereas the integrability of the
nearest neighbour Hubbard model follows from its connection to
an integrable system of two coupled six-vertex models
\cite{Shastry88a}. We expect that a proof of the integrability of
the non-nearest-neighbour Hubbard models will reveal a more
generic structure.

A first application of our new Y($sl_2$) Yangian will be the
classification of the Jastrow-like eigenfunctions of the ``$t$-0''
model \cite{WZC93}. For the system at finite on-site energy the
situation is more complicated. Not even the ground state wave
function is known. There is evidence that the wave functions are
neither of Jastrow-type as for the Haldane-Shastry chain nor of
Bethe Ansatz form as in case of the nearest neighbour Hubbard
model \cite{BaGe95}.

{\bf Acknowledgments}. This  work has been supported by the Japan
Society for the Promotion of Science and the Ministry of Science,
Culture and Education of Japan. One of the authors (V. I.) would
like to express his sincere gratitude to Professor Minoru Takahashi
for useful discussions and kind hospitality extended to him at
the ISSP. The other one (F. G.) likes to thank Professor Miki
Wadati for his hospitality and the warm atmosphere in his group
at Tokyo University.

\end{document}